\documentclass[sigconf]{acmart}

\usepackage{algorithmic}
\usepackage{graphicx}
\usepackage{textcomp}
\usepackage{xcolor}
\usepackage{url}
\usepackage{listings}
\usepackage{capt-of}
\usepackage{caption}

\lstdefinestyle{prompt}{
  basicstyle=\small\ttfamily,
  breaklines=true,
  columns=fullflexible,
  keepspaces=true,
  frame=single
}

\definecolor{codekw}{RGB}{0,0,150}
\definecolor{codeid}{RGB}{38,38,38}
\definecolor{codecm}{RGB}{0,128,0}
\definecolor{codestr}{RGB}{163,21,21}
\definecolor{codenums}{gray}{0.45}

\lstdefinelanguage{JavaScript}{
  morekeywords={typeof,new,true,false,catch,function,return,null,switch,var,if,in,while,do,else,case,break,for,let,const,await,async,
                class,export,import,this}, % podes deixar aqui; não interfere
  sensitive=true,
  comment=[l]{//},
  morecomment=[s]{/*}{*/},
  morestring=[b]', morestring=[b]"
}

\lstdefinelanguage{TypeScript}{
  language=JavaScript,
  keywords={interface,type,implements,private,public,protected,readonly,enum,extends,super,this},
  morekeywords=[2]{import,export,from,as}
}

\lstdefinestyle{ts}{
  language=TypeScript,
  basicstyle=\footnotesize\ttfamily,
  keywordstyle=\color{codekw}\bfseries,
  keywordstyle=[2]\color{blue}\bfseries,
  basicstyle=\footnotesize\ttfamily,
  keywordstyle=\color{codekw}\bfseries,
  ndkeywordstyle=\color{codekw},
  commentstyle=\color{codecm}\itshape,
  stringstyle=\color{codestr},
  identifierstyle=\color{codeid},
  numbers=left,
  numberstyle=\tiny\color{codenums},
  numbersep=8pt,
  columns=fullflexible,
  keepspaces=true,
  breaklines=true,
  showstringspaces=false,
  tabsize=2
}

\AtBeginDocument{%
  }

\copyrightyear{2026}
\acmYear{2026}
\setcopyright{cc}
\setcctype{by}
\acmConference[AST '26]{7th ACM/IEEE International Conference on Automation of Software Test (AST 2026)}{April 13--14, 2026}{Rio de Janeiro, Brazil}
\acmBooktitle{7th ACM/IEEE International Conference on Automation of Software Test (AST 2026) (AST '26), April 13--14, 2026, Rio de Janeiro, Brazil}
\acmPrice{}
\acmDOI{10.1145/3793654.3793743}
\acmISBN{979-8-4007-2476-3/2026/04}

\begin{document}

\title{APITestGenie: Generating Web API Tests from Requirements and API Specifications with LLMs}

\author{André Pereira}
\affiliation{%
  \institution{Deloitte and Faculty of Engineering, University of Porto}
  \city{Porto}
  \country{Portugal}}
\email{adbp@live.com.pt}

\author{Bruno Lima}
\affiliation{%
  \institution{LIACC, Faculty of Engineering, University of Porto}
  \city{Porto}
  \country{Portugal}}
\email{brunolima@fe.up.pt}

\author{João Pascoal Faria}
\affiliation{%
  \institution{INESC TEC, Faculty of Engineering, University of Porto}
  \city{Porto}
  \country{Portugal}}
\email{jpf@fe.up.pt}


\begin{abstract}
Modern software systems rely heavily on Web APIs, yet creating meaningful and executable test scripts remains a largely manual, time-consuming, and error-prone task. In this paper, we present \textit{APITestGenie}, a novel tool that leverages Large Language Models (LLMs), Retrieval-Augmented Generation (RAG), and prompt engineering to automatically generate API integration tests directly from business requirements and OpenAPI specifications.
We evaluated \textit{APITestGenie} on 10 real-world APIs, including 8 APIs comprising circa 1,000 live endpoints from an industrial partner in the automotive domain. The tool was able to generate syntactically and semantically valid test scripts for 89\% of the business requirements under test after at most three attempts. Notably, some generated tests revealed previously unknown defects in the APIs, including integration issues between endpoints.
Statistical analysis identified API complexity and level of detail in business requirements as primary factors influencing success rates, with the level of detail in API documentation also affecting outcomes.
Feedback from industry practitioners confirmed strong interest in adoption, substantially reducing the manual effort in writing acceptance tests, and improving the alignment between tests and business requirements.
\end{abstract}

\begin{CCSXML}
<ccs2012>
   <concept>
       <concept_id>10011007.10011074.10011099</concept_id>
       <concept_desc>Software and its engineering~Software verification and validation</concept_desc>
       <concept_significance>500</concept_significance>
       </concept>
 </ccs2012>
\end{CCSXML}

\ccsdesc[500]{Software and its engineering~Software verification and validation}

\keywords{API specifications, API testing, Business requirements,  
Large language models, OpenAPI, Test script generation}

%\received{20 February 2025}

\maketitle

\section{Introduction}
\label{sec1}

Generative Artificial Intelligence (AI) has witnessed remarkable progress in recent years, reshaping the landscape of intelligent systems with applications in many domains. Studies show that intelligent assistants powered by Large Language Models (LLMs) can now generate program and test code with high accuracy based on textual prompts~\cite{ref:copilotSimpleProblems}, boosting developers' and testers' productivity. 

Although LLMs have been explored for several test generation tasks, there is a lack of studies exploring them for testing Web APIs, which constitute a fundamental building block of modern software systems. 
The rapid proliferation of APIs in recent years~\cite{ref:stateAPI2023} has further heightened the need for scalable and automated testing approaches that can keep pace with rapid software evolution and deployment cycles.

Identifying and creating relevant API tests is a tedious, time-consuming task~\cite{ref:testAutomation}, requiring significant programming expertise, particularly as system complexity increases.
While integration and system tests ensure functional integrity, they are more challenging to generate because the expected behavior is often described at a high-level of abstraction. 
 Existing API test automation tools often fail to leverage high-level requirements to produce comprehensive test cases and detect semantic faults.

To overcome such limitations by taking advantage of recent developments in Generative AI, we present \textit{APITestGenie} -- a tool that leverages LLMs to generate executable API test scripts from business requirements written in natural language and API specifications documented using OpenAPI~\cite{ref:openAPIonline}, the most widely adopted standard for documenting API services. 
APITestGenie was developed and validated in collaboration with an industrial partner in the automotive domain.
%, as part of a broader research initiative on intelligent test automation.

We evaluated \textit{APITestGenie} on 10 real-world APIs, including 8 APIs comprising circa 1,000 live endpoints from our industrial partner. The tool was able to generate syntactically and semantically valid test scripts (requiring no manual editing) for 89\% of the business requirements under test after at most three attempts.

To summarize, the main contributions of this paper are:
\begin{itemize}
    \item A novel tool (\textit{APITestGenie}) that leverages LLMs, retrieval-augmented generation (RAG), and prompt engineering to generate executable API test scripts from business requirements in natural language and OpenAPI specifications of Web APIs.
    \item Results of an experimental evaluation of the tool on APIs of varying complexity from an industrial partner in the automotive domain, obtaining up to 89\% valid test scripts, a significant reduction in manual effort, and evidence of the influence of API complexity and requirements detail on test generation success.
\end{itemize}

In next sections, we analyze related work, present our solution and its evaluation, and point out conclusions and future work.

\section{Related Work}
\label{sec2}

Developers often perceive testing as a high-effort, low-reward activity. Empirical studies show that even when developers spend up to 40\% of their time on testing, it is frequently seen as excessive given the perceived complexity and limited immediate value~\cite{surveydevelopersthinktesting}. This effort is further undermined by the lack of recognition that testing receives within teams, leading developers to prioritize feature delivery over comprehensive test coverage. These factors contribute to insufficient testing practices in many software projects and motivate the need for automation tools that reduce the manual burden of test creation while increasing the value of testing within development workflows.

In recent years, several approaches and tools have emerged for testing REST APIs, with a significant rise in research since 2017, focusing on test automation using black-box techniques and OpenAPI schemas~\cite{golmohammadi2023testing}. Some recent approaches also take advantage of LLMs.

In a 2022 study~\cite{kim2022automated}, the authors compared the performance of ten state-of-the-art REST API testing tools from both industry and academia on a benchmark of twenty open-source APIs, in terms of code coverage and failures triggered (5xx status codes), highlighting EvoMaster among the best-performing.

EvoMaster~\cite{ref:evoMasterOnline} can generate test cases for REST APIs in both white-box and black-box modes, with worse results in the former due to the lack of code analysis. It works by evolving test cases from an initial random population using an evolutionary algorithm, trying to maximize measures such as code coverage (only in white-sbox mode) and fault detection, using several heuristics~\cite{ref:apiAutomatedTestGeneration}. Potential faults considered for fault finding are based on HTTP responses with 5xx status codes and discrepancies between API responses and expectations derived from OpenAPI schemas.

However, like other approaches that rely solely on OpenAPI specifications as input~\cite{ref:quickRest,ref:gpt_sbst,ref:apiBlackBoxTesting}, EvoMaster is limited by the absence of requirements specifications and context awareness, which restricts the variety of tests that can be generated and the types of faults that can be detected. Our goal is to leverage requirements specifications and the context-awareness capabilities of LLMs to overcome these limitations, enabling the generation of more comprehensive and effective tests.

Recent works have started leveraging LLMs for REST API testing. 
RESTGPT~\cite{kim2024leveraging} enhances testing by extracting rules and parameter values from OpenAPI descriptions but does not generate executable scripts. RESTSpecIT~\cite{decrop2024you} uses LLMs for automated specification inference, discovering undocumented routes and query parameters, and identifying server errors.
Both approaches complement ours, as the enhanced or inferred specifications can be used as input for subsequent test script generation.

LlamaRestTest~\cite{kim2025llamaresttest} employs fine-tuned and quantized language models to infer inter-parameter dependencies and dynamically generate realistic inputs based on server feedback, achieving high code coverage. Its strength lies in adaptability during runtime, though it primarily relies on API specifications without explicit linkage to stakeholder requirements.

AutoRestTest~\cite{stennett2025autoresttest} adopts a multi-agent reinforcement learning framework enriched with Semantic Property Dependency Graphs (SPDG), integrating LLM-driven inputs to intelligently manage operation ordering and parameter values. This approach significantly improves fault detection and test coverage, yet it still does not incorporate natural-language business requirements into its input process.

LogiAgent~\cite{zhang2025logiagent} advances beyond traditional status-code validation by introducing logical oracles through a multi-agent LLM framework. By verifying the logical correctness of responses against expected business behaviors, LogiAgent uncovers subtle domain-specific faults. However, its generation of test scenarios is based only on the API specification and does not explicitly utilize natural-language business requirements as input.

DeepREST~\cite{corradini2024deeprest} applies curiosity-driven deep reinforcement learning to implicitly uncover API constraints and execution orders. Despite its capability to expose unforeseen faults and improve coverage, it does not integrate textual stakeholder requirements or leverage natural-language contexts provided by end-users.

In contrast, our proposed approach, \textit{APITestGenie}, uniquely distinguishes itself by explicitly incorporating natural-language business requirements alongside OpenAPI specifications through Re-trieval-Augmented Generation (RAG). Thus, \textit{APITestGenie} bridges the critical gap between user-centric business expectations and automated REST API testing, complementing other approaches.

\section{Solution Design}
\label{sec:solution}

\subsection{Architecture and Workflow}

\begin{figure}[h]
\newlength{\xfigwd}
\setlength{\xfigwd}{\linewidth}
\centerline{\includegraphics[width=21pc]{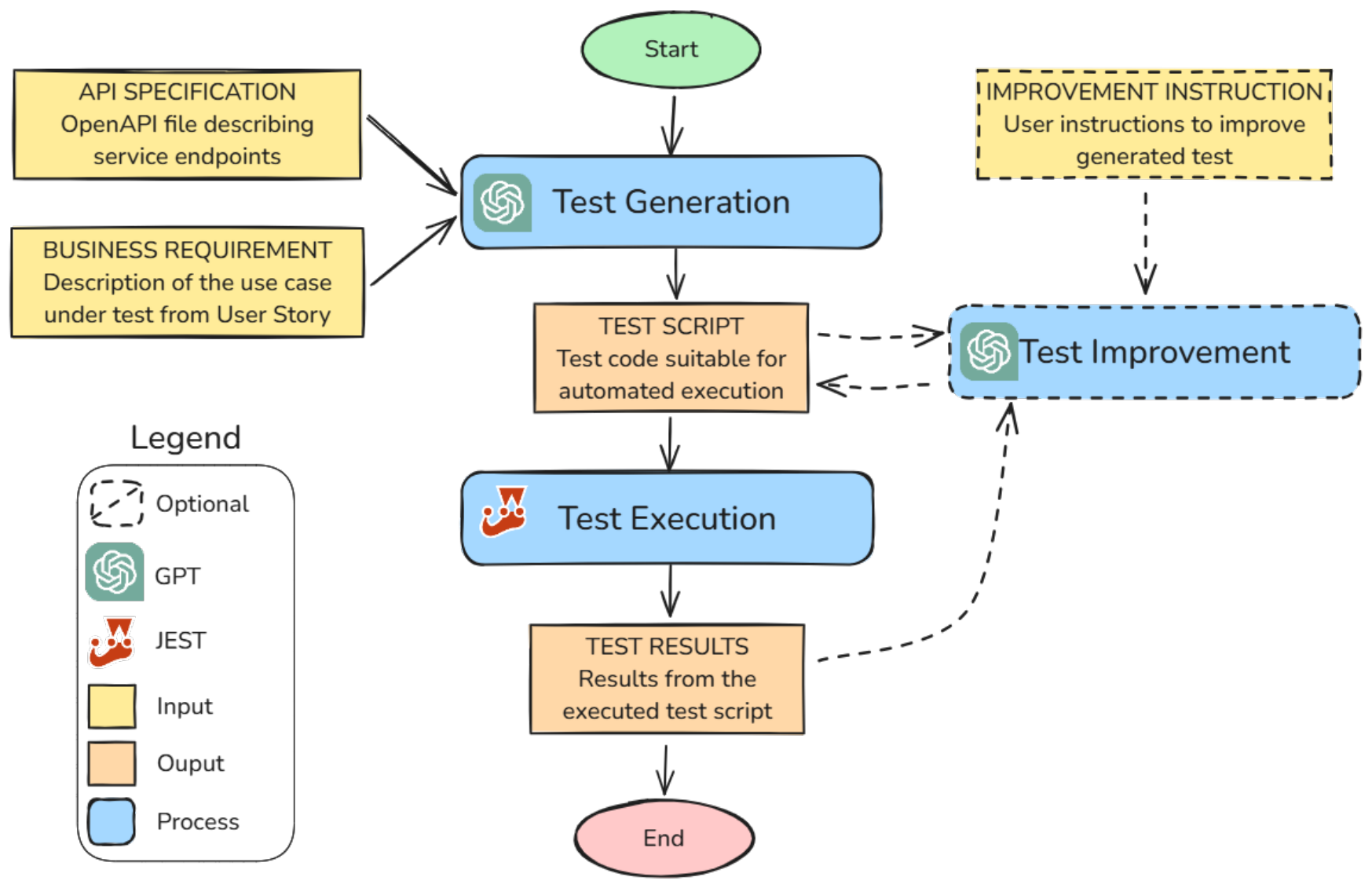}}
\caption{APITestGenie flow diagram, showcasing the main processes in the system, inputs and outputs.}
\label{fig:userArchitecture}
\vspace*{-10pt}
\end{figure}

APITestGenie autonomously generates, improves, and executes test cases. Its architecture (see Figure~\ref{fig:userArchitecture}) is divided into three modular processes: \textit{Test Generation}, \textit{Test Improvement}, and \textit{Test Execution}. While each flow can operate independently, they are designed to work together for a more complete solution.

The \textit{Test Generation} process receives as input the API specification and the business requirement to generate a test script. 
Depending on the API size, we use either retrieval-augmented generation (RAG) for larger API specs or full processing for smaller ones. We then construct the system and user prompts based on the TELeR taxonomy~\cite{ref:teler} for a \textit{$<$single turn, instruction, high detail, defined$>$} use case (see Section~\ref{sec:sys_user_prompt}). 
Lastly, the LLM is invoked and the generated content is parsed, producing the final executable test script. 
An example of a generated test script is included in the Appendix.

The \textit{Test Improvement} process can be used to refine the generated script based on previous test results and user-provided feedback.

The \textit{Test Execution} process uses the Jest framework in a TypeScript environment to execute the generated test script and report the results.

The design promotes modularity, facilitation the integration or upgrading of components.The clear separation of responsibilities among components enhances their specialization and effectiveness, ultimately improving the quality and accuracy of generated content.

\vspace*{-5pt}
\subsection{System and User Prompt}
\label{sec:sys_user_prompt}

The system prompt is structured as follows:
\begin{enumerate}
    \item \textit{Context} — specifies the task to be performed (see Figure~\ref{fig:systemPromptContext}), including:
    \begin{enumerate}
        \item Model's role;
        \item High-level objetive;
%        \item Target user and objective guidelines;
        \item Test structure example;
        \item Information on available environment variables.
    \end{enumerate}
    \item \textit{Performance} — guidelines to evaluate the LLM output (see Figure~\ref{fig:systemPromptPerformance}), including:
    \begin{enumerate}
        \item Output evaluation metrics;
        \item Generation guidelines.
    \end{enumerate}
    \item \textit{Output} — describes the reasoning steps to generate the output (see Figure~\ref{fig:systemPromptOutputStrucuture}):
    \begin{enumerate}
        \item Reasoning steps;
        \item Output components and structure.
    \end{enumerate}
\end{enumerate}

\vspace{8pt}
The user prompt is structured as follows:
\begin{enumerate}
    \item \textit{Requirement} — business requirement under test (usually written in the format of a user story);
    \item \textit{API} — API specification under test (in OpenAPI).
\end{enumerate}

\vspace{10 pt}
%\begin{minipage}{\linewidth}
\begin{lstlisting}[style=prompt]
As an AI coding assistant, my goal is to facilitate the creation of executable API integration tests in TypeScript.
Many users may not be familiar with coding, so I am here to bridge the gap and help them craft tests that validate their application's business requirements.

To ensure the tests are practical and meet the users' needs, I will generate integration test cases in TypeScript using the Axios and Jest libraries.
These tests will be designed for immediate execution and will interact with the API endpoints defined in the API specification.

I am expected to create more complex tests following the example
***
{test_example}
***
    \end{lstlisting}
    \vspace{-\abovecaptionskip}
    \captionof{figure}{``Context'' section of the system prompt.}
%    , defining the model's role, its high-level objective, a sample test structure and details about accessible environment variables.}
    \label{fig:systemPromptContext}
%\end{minipage}

\vspace{8pt}

%\begin{minipage}{\linewidth}
\begin{lstlisting}[style=prompt]
TESTS WILL BE ASSESSED ON SEVERAL KEY FACTORS:
- Executability: The tests must run smoothly and without errors.
- Relevance: The tests must be meaningful and appropriate for the requirement.
- Correctness: The test code should be free of logical errors.
- Coverage: The tests should cover as many endpoints and scenarios as possible to ensure thorough validation.
- Code Quality: The code is reliable, executable, and includes clear explanations where necessary.
- Endpoint Accuracy: The API call matches the type of the request and response object. 

GUIDELINES FOR TEST GENERATION:
- Informational Completeness: Collect as much information as possible. Additionally, generate experimental data to use in placeholders.
- Environment Setup: Ensure all necessary data is collected to guarantee that the test can execute correctly.
    \end{lstlisting}
    \vspace{-\abovecaptionskip}
    \captionof{figure}{``Performance'' section of the system prompt.}
%    defining the evaluation guidelines for LLM output.}
    \label{fig:systemPromptPerformance} 
%\end{minipage}

\vspace{10pt}

%\begin{minipage}{\linewidth}
\begin{lstlisting}[style=prompt]
Here's how the test generation will be accomplished:
\\
1. CLARIFYING THE BUSINESS REQUIREMENT:
- Summarize the business requirement and explain what aspects the tests verify.
- Describe the data and environment state of the test.

2. LISTING ENDPOINTS:
- List the endpoints that the test script will interact with. 
- For each endpoint, specify the types of the request and response objects.
- Provide a brief list of steps to reach the correct test environment state.

3. CRAFT EXECUTABLE TEST CODE:
- Generate test cases in TypeScript using Axios and Jest. 
- Present the code in a single block, ready to run.
- Explain unclear properties and minimize extraneous prose.

Document your generation using the following format:

REQUIREMENT:
<1. **Clarifying the Business Requirement**>
ENDPOINTS:
<2. **Listing Endpoints**>
TEST:
```typescript
<3. **Craft Executable Test Code**>

\end{lstlisting}
\vspace{-\abovecaptionskip}
    \captionof{figure}{``Output'' section of the system prompt.}
%    , describing the expected reasoning steps and final output structure.}
\label{fig:systemPromptOutputStrucuture} 
%\end{minipage}

\indent
\setlength{\parindent}{1em}

\newpage

\subsection{API Pre-processing and Retrieval-Augmented Generation}

A significant limitation of current LLM models is their context window size.
To mitigate this, we first created a preprocessing script that simplifies the raw OpenAPI specification by removing image tags and administrative or deprecated resources. This approach reduces token usage while keeping the essential content. Although effective for smaller APIs, this method proved insufficient for larger ones, motivating the adoption of a more robust technique.

%\subsection{Prompt with Retrieval-Augmented Generation (RAG)}

To overcome the limitations of LLM context windows when dealing with large API specifications, we implemented a Retrieval-Augmented Generation (RAG)~\cite{ref:metaRAG} pipeline. This process allows selective retrieval of the most relevant parts of an API specification to include in the prompt provided to the LLM. The pipeline begins by preprocessing the OpenAPI specification and splitting it into semantically coherent chunks ranging from 800 to 1200 tokens, preserving contextual consistency while maintaining conciseness. Each chunk is then embedded using a text-embedding model, and the resulting vectors, along with their documents, are stored in a vector database.

To retrieve relevant content for a given business requirement, we use a multi-step query process. The requirement is first expanded using an LLM-based script that rephrases it into multiple semantically similar variants, capturing different perspectives of the same intent. Each expanded version is embedded using the same model and used to query the vector database. The retrieved chunks are then aggregated to form a consolidated set of relevant documents, which are appended to the system prompt for test generation. Because retrieval is based on multiple semantically equivalent versions, the final context is robust and less susceptible to hallucinations.

\begin{figure}[h]
\setlength{\xfigwd}{\linewidth}
\centerline{\includegraphics[width=20pc]{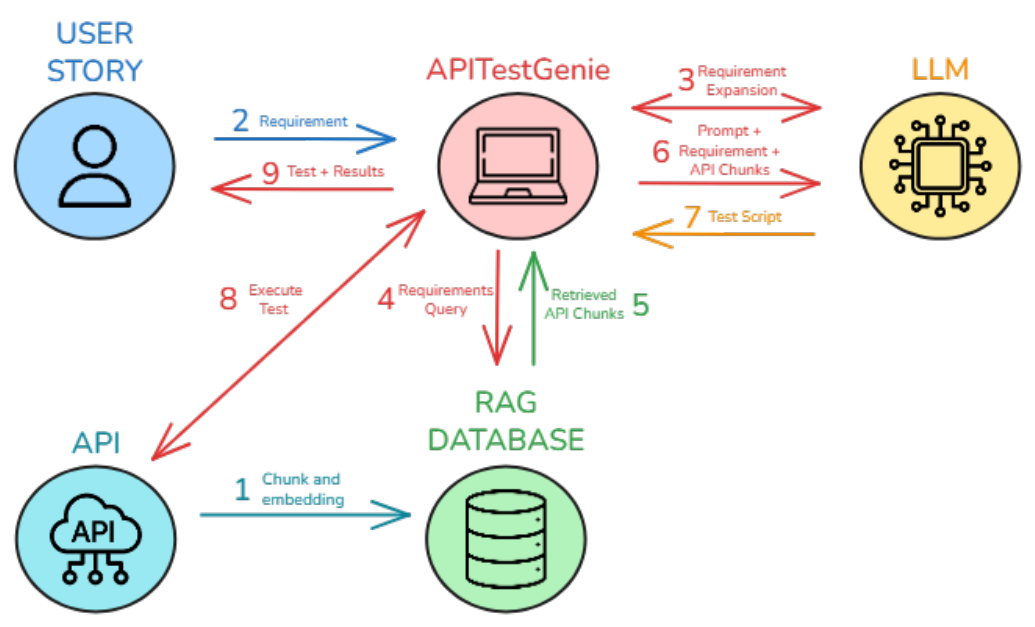}}
\caption{APITestGenie’s RAG process builds a database from the API specification, expands requirements, and retrieves relevant API segments to give the LLM precise context for generating targeted test scripts.}
\label{fig:ragArchitecture}
\vspace*{-10pt}
\end{figure}

This RAG-enhanced prompt construction enables \textit{APITestGenie} to scale effectively to large and complex APIs without exceeding the LLM’s input limitations, while maintaining alignment with the business logic expressed in user requirements. For a visual representation of the flow with RAG, see Figure~\ref{fig:ragArchitecture}.

\section{Evaluation}
\label{sec:evaluation}

\begin{table*}[h]
    \caption{Experimental test generation results for 10 APIs and 25 business requirements.}
    \centering
    {\fontsize{9}{13}\selectfont
    \begin{tabular}{|l|c|c|c|c|c|c|c|}
        \hline
        \textbf{API} & \textbf{Access} & \textbf{Original} & \textbf{Simplified}  & \textbf{Business} & \textbf{Valid Test} & \textbf{BR with} & \textbf{Generation}\\
        &  & \textbf{tokens$^a$} & \textbf{tokens$^b$} & \textbf{Req. (BR)}& \textbf{Scripts (TS)$^c$} & \textbf{$\ge$1 valid TS} & \textbf{Time (avg.)} \\ 
        \hline
        CF.1 Cat Fact$^e$ & Public & 754 & 754  & 1  & 2 & 1 & 92s \\
        PS.1 Pet Shop$^e$ &  Public & 4,070 & 4,070  & 2 & 4 & 2 & 75s \\
        %UCP Otd (L1) 
        V.2 Vehicle Configuration& Private & 430,739 & 16,094 & 5  & 15 & 5 & 102s \\
        \textbf{Subtotal Low API Complexity} &  &  & & \textbf{8}  & \textbf{21 (87.5\%)} & \textbf{8 (100\%)} &  \\
        \hline
        
        %GCDM Customer (L2)
        C.1 Customer Authorization&  Private & 25,966 & 25,475  & 2 & 5 & 2 & 113s \\
        %GCDM (L2) 
        C.2 Customer Account& Private & 47,497 & 47,004  & 2 & 5 & 2 & 98s \\
        %OneShop (L3) 
        S.1 Web store& Private & 49,761 & 49,761 & 2  & 3 & 1& 154s \\
        %UCP Accessories (L2) 
        V.1 Vehicle Accessories& Private & 882,272 & 14,018 & 2  & 3 & 1 & 115s \\
        %UCP Localisations (L2) 
        V.3 Vehicle Localization& Private & 1,620,488 & 82,267 & 2  & 3& 2 & 131s \\
        %UCP Vehicles (L2) 
        V.4 Vehicle Production& Private & 417,109 & 108,724$^f$ &4  & 8 & 4 & 146s \\
        %UCP All (L2) 
        V.5 All Vehicle Services & Private &  5,021,555 & 424,465$^f$ &  3 & 4 & 2 & 227s \\ 
        \textbf{Subtotal High API Complexity} &  &  &  & \textbf{17} & \textbf{31 (60.8\%)} & \textbf{14 (82.4\%)} & \\
        \hline
        \textbf{Total} & & \textbf{9,254,211} & \textbf{772,632(8.3\%)} &  \textbf{25}  &  \textbf{52 (69.3\%)}& \textbf{22} \textbf{(88.6\%)}& \textbf{126s} \\
        \hline
        \multicolumn{8}{l}{$^{\mathrm{a}}$Number of tokens in the original OpenAPI specification counted with the tiktoken tokenizer.} \\ 
        \multicolumn{8}{l}{$^{\mathrm{b}}$Number of tokens in the simplified spec (used as input for test generation), after removing irrelevant tags and resources.} \\ 
    %    \multicolumn{8}{l}{$^{\mathrm{d}}$Number of business requirements tested.} \\
        \multicolumn{8}{l}{$^{\mathrm{c}}$Number of syntactically and semantically valid test scripts in three generation attempts per business requirement.} \\
        \multicolumn{8}{l}{$^{\mathrm{d}}$RAG was used to filter simplified API specs with over 100K tokens, reducing them to fit the LLM's maximum prompt size.} \\
        \multicolumn{8}{l}{$^{\mathrm{e}}$Public APIs available at \url{https://catfact.ninja/} and  \url{https://petstore.swagger.io/}.} \\
    \end{tabular}
    }
    \label{tab:generationByService}
\end{table*}

We evaluated \textit{APITestGenie} in two phases. First, we conducted an experimental evaluation on ten APIs (including eight from our industrial partner in the automotive domain), collecting metrics on correctness, generation time, and cost of the generated test scripts. Second, we organized a practical workshop with fifteen experts from our industrial partner to gather feedback on the tool’s capabilities and adoption.
We also compared \textit{APITestGenie} with EvoMaster (a state-of-the-art tool) for one of the public APIs tested.

The main research questions we want to answer are:
\begin{itemize}
  \item \textbf{RQ1} — How effectively and efficiently can \textit{APITestGenie} generate API test scripts from business requirements and API specifications?
  \item \textbf{RQ2} — How do API complexity, API documentation detail and requirements detail influence the effectiveness of test script generation by \textit{APITestGenie}?
  \item \textbf{RQ3} — How do industry practitioners perceive the usefulness and adoptability of \textit{APITestGenie} for API test script generation?
\end{itemize}

\subsection{Experimental Evaluation}

\subsubsection{Materials and methods}

We assessed the robustness of test scripts generated by \textit{APITestGenie} for ten APIs with varying complexity and documentation levels (see Table~\ref{tab:generationByService}). 

\begin{figure}[h]
\centerline{\includegraphics[width=0.85\columnwidth,trim={0 0 6.9cm 0},clip]{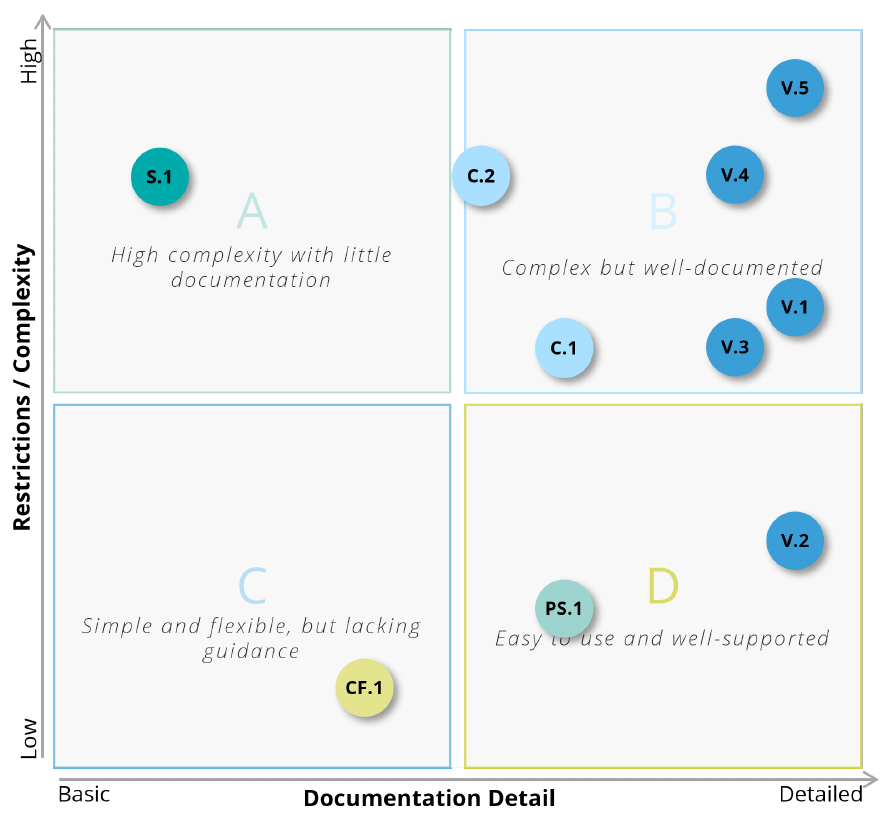}}
\caption{Characterization of the APIs under test according to their internal complexity and documentation detail.}
\label{fig:apiMatrix}
\vspace*{-10pt}
\end{figure}

\vspace{4pt}

To help answer RQ2, we classified the APIs under test along two dimensions (see Figure~\ref{fig:apiMatrix}): 
\begin{itemize}
\item Documentation detail – considering whether API methods are properly documented, examples of calls are provided, possible errors are listed, and examples of input data (parameters) and expected outputs (response schemas) are included;
\item API complexity – considering the number of available endpoints, the relationships and dependencies among them, the presence of authentication requirements, and the predictability of API responses and errors.
\end{itemize}

\noindent
\setlength{\parindent}{0em}

\begin{minipage}{\linewidth}
\begin{lstlisting}[style=prompt]
1. Basic BR description (user story): 
As a user, I want to receive a new and random cat fact every time I open the app or refresh the content so that I can learn intriguing information about cats and stay engaged with the app.

2. BR including concrete data: 
As a user, I want to retrieve all available models in Germany so that I can know which products are available.

3. BR including procedural information: 
As a user, I want to retrieve the price of accessories for a product in a given market. First, use the product service to get a list of all available models. Then, use the accessories service to retrieve available accessories, and finally use the pricing service to calculate their prices.
\end{lstlisting}
\vspace{-\abovecaptionskip}
    \captionof{figure}{Examples of business requirements (BR) with varying detail.}
\label{fig:brdetail} 
\end{minipage}

\vspace{10 pt}

\indent
\setlength{\parindent}{1em}

Three of the APIs are low-complexity, with two of them being public APIs (CF.1 Cat Fact and PS.1 Pet Shop) and the other private. 
The other seven private APIs from our industrial partner in the automotive domain have a higher level of complexity, comprising approximately 1,000 live endpoints.

Overall, we used 25 unique business requirements (BR) 
%with varying levels of detail (L1 to L3) 
as input for test script (TS) generation. The BR were usually described in the form of user stories, with varying levels of detail (see Figure~\ref{fig:brdetail}).

To improve statistical significance and minimize the effect of LLM randomness on the results, we performed three independent test script generation attempts per BR, totalling 75 generation attempts. 

The model used in the experiment was GPT-4-Turbo, featuring a 128k-token context window, deployed under an enterprise configuration with data privacy safeguards. 

We then executed the generated test scripts to check their syntactic validity and manually inspected the executable ones to determine their semantic validity.  
By a \emph{valid test script}, we mean a syntactically and semantically valid test script, i.e.,  a test script that executes without syntax errors and is considered semantically accurate in manual inspection (confirming or refuting the business requirement when run against the target API). 

\subsubsection{Overall performance results}

As shown by the totals in Table~\ref{tab:generationByService}, 52 out of 75 generation attempts (three per business requirement) resulted in valid test scripts, corresponding to an overall success rate of 69.3\%. 
In total, 88.6\% of the business requirements had at least one valid test script generated within three attempts.

% \textcolor{red}{Since syntactically invalid or empty results (4/75) can be automatically detected and regenerated, a more practical success rate is the ratio between valid scripts and scripts requiring manual semantic validation: 52 out of 71 --- 73.2\% valid executable test scripts.}      

% \textcolor{red}{The 75 scripts contained 90 test cases, 71 of which were valid --- 78.9\% valid test cases.}

Average time and cost were 126s and 0.37€ per generation, respectively, with a tendency to decrease as newer LLMs are adopted. 

These results indicate that \textit{APITestGenie} can effectively generate executable and functionally meaningful test scripts from business requirements and API specifications.

\subsubsection{Assessment of the generated test scripts}

Table~\ref{tab:generationAssessment} presents the results of the assessment of the generated test scripts.

\begin{table*}[h]
    \caption{Assessment of the generated test scripts, distinguishing valid and invalid test scripts.}
    \centering
    {\fontsize{9}{12}\selectfont
    \begin{tabular}{|p{3.1cm}|c|p{8cm}|p{3.5cm}|}
        \hline
        \textbf{Result Type} & \textbf{Test Scripts} & \textbf{Description} & \textbf{Resolution} \\ 
        \hline
        \hline
        \textbf{Invalid:} & \textbf{23 (30.7\%)} & & \\
        \hline
        a) Empty script $^a$ & 1 & The generated script is empty. & Re-generate. \\
        \hline
        b) Syntax error $^a$ & 3 & The script cannot be executed due to syntax errors. & Re-generate. \\
        \hline
        c) Semantic error(s) & 19 & Test execution failed due to LLM hallucinations: wrong/non-existent import, attribute, operation, response structure validation, etc. & Fix script, re-generate, or run improvement flow (including error information). \\
        \hline
        \hline
        \textbf{Valid:} & \textbf{52 (69.3\%)} & & \\
        \hline
        d) Pass  & 31 & Test execution succeeds, and the script is considered semantically valid in manual inspection.$^b$ & Save script for regression testing. \\
\hline
        % Fail: API defect & 12 & Test execution fails due to a defect in the API specification or implementation, making it inconsistent with the business requirements. Failures include inconsistencies with requirements in tests involving multiple endpoints. & Fix specification/implementation and re-run. \\

        e) Fail: API defect & 12 & 
Test execution fails due to inconsistencies between the API specification or implementation and the expected behavior described in the requirements. 
Failures include inconsistencies with requirements in tests involving multiple endpoints. & 
Fix specification or implementation and re-run. \\

\hline
        f) Fail: Insufficient documentation & 2 & Test execution fails because key information is missing in the API specification. & Fix specification and re-generate. \\ 
\hline
        g) Fail: Environment issue & 7 & Test execution fails due to missing or incorrect environment variables (e.g., access tokens). & Fix environment setup and re-generate. \\      
        \hline
        \hline
        \multicolumn{4}{l}{$^{\mathrm{a}}$These errors are autonomously identified (non-executable), enabling automatic reattempt.} \\
        \multicolumn{4}{l}{$^{\mathrm{b}}$In our experiment, all scripts that executed successfully and passed were considered valid in manual inspection.} \\
    \end{tabular}
    }
    \label{tab:generationAssessment}
\end{table*}

Out of 75 generation attempts, 4 (5.3\%, $a$ and $b$) resulted in empty or syntactically invalid scripts, which can be automatically identified and typically resolved by regeneration.    

The remaining scripts were executed, their test results collected, and the scripts manually inspected for semantic validity. 

In 19 cases (25.3\%, $c$), test execution failed due to semantic errors. In practice, such cases of LLM hallucination can be resolved by minor fixes, regeneration, or by including error information in the improvement prompt. 

In 31 cases (41.3\%, $d$), test execution succeeded and the scripts were considered semantically valid.

The remaining 21 cases (28\%) corresponded to semantically valid scripts whose execution failed due to API defects (12 cases, 16\%), insufficient API documentation (2 cases, 3\%), or environment issues (7 cases, 9\%). 

These results highlight the defect-detection capability of the generated tests (19\% revealed problems in the API implementation or documentation) and the potential to reduce manual (re)writing effort (only 19 out of the 71 inspected scripts might require minor fixes). 

The Appendix includes an example script that interacts with two endpoints and uncovers an API defect that would be difficult to detect with unit tests.

\subsubsection{Analysis of influencing factors}

We investigated whether API complexity, API documentation detail, and business requirements detail had a statistically significant impact on the outcomes. For each group, we compared the number of valid scripts to the total number of generation attempts (three per requirement) using the chi-squared test.

API complexity showed a statistically significant effect. Low-complexity APIs had a substantially higher proportion of valid scripts (87.5\% in Table~\ref{tab:generationByService}) than high-complexity APIs (60.8\% in Table~\ref{tab:generationByService}), with a p-value of 0.038 in the chi-squared test. In contrast, documentation detail (p = 0.57) did not show a significant effect.

Regarding the impact of the level of detail in the business requirements, we observed that requirements including concrete data (6 BRs) led to a higher proportion of valid test scripts (88.9\%) compared to those without concrete data (19 BRs, 63.2\% success rate). The impact on test generation success rate was statistically significant, with a p-value of 0.039 in the chi-squared test. The inclusion of procedural information did not show a statistically significant impact.   

These findings suggest that, while improving documentation may help, reducing API complexity (e.g., by decomposing large APIs into smaller services) and providing concrete data in the business requirements had the greatest measurable impact in our experiments.

\subsection{Comparison with EvoMaster}

To contextualize the performance of APITestGenie, we compared it with EvoMaster~\cite{ref:evoMasterOnline}, one of the most established tools for automated REST API test generation. 
EvoMaster was executed in black-box mode against the same Pet Store API used in our example test, see Appendix, Listing~\ref{lst:testScript}. 
Both tools were limited to a 70~s execution/generation window, which corresponds approximately to the average time required for APITestGenie to produce a test script.

EvoMaster generated 27 test cases in total. Of these, \textbf{22 (81\%)} triggered failures, \textbf{2 (7\%)} executed successfully, and \textbf{3 (11\%)} required manual validation. 
According to EvoMaster’s internal metrics, 19 of the reported failures were due to structural mismatches between the received responses and the OpenAPI schema 
\textit{"Received a Response From API With a Structure/Data That Is Not Matching Its Schema"}, and one failure corresponded to \textit{"HTTP Status 500"}. 
Under our evaluation framework, such schema-mismatch errors would be classified as documentation defects, while the 500 status case, caused by malformed input data, would be considered an invalid test case resulting from a hallucination.

From a functional perspective, EvoMaster successfully detects schema and documentation inconsistencies but focuses on testing each endpoint independently. 
In contrast, APITestGenie generates multi-endpoint integration tests directly derived from business requirements. 
For the Pet Store scenario, APITestGenie produced a two-step test flow that (i) creates a pet using \texttt{POST /v2/pet} and (ii) retrieves it using \texttt{GET /v2/pet/\{petId\}}. 
The generated test revealed that the retrieved pet differed from the one created, an integration-level defect that EvoMaster, which treats endpoints in isolation, cannot detect.

In summary, EvoMaster achieved endpoint-level structural validation and uncovered documentation defects efficiently, whereas APITestGenie provided requirement-driven, end-to-end validation capable of exposing logical or cross-endpoint faults. 
These results highlight the complementary nature of the two tools: EvoMaster excels at structural conformance and schema testing, while APITestGenie extends coverage to business-level behaviors and system integration flows.

\subsection{User Feedback}

\begin{figure}
\centerline{\includegraphics[width=21pc]{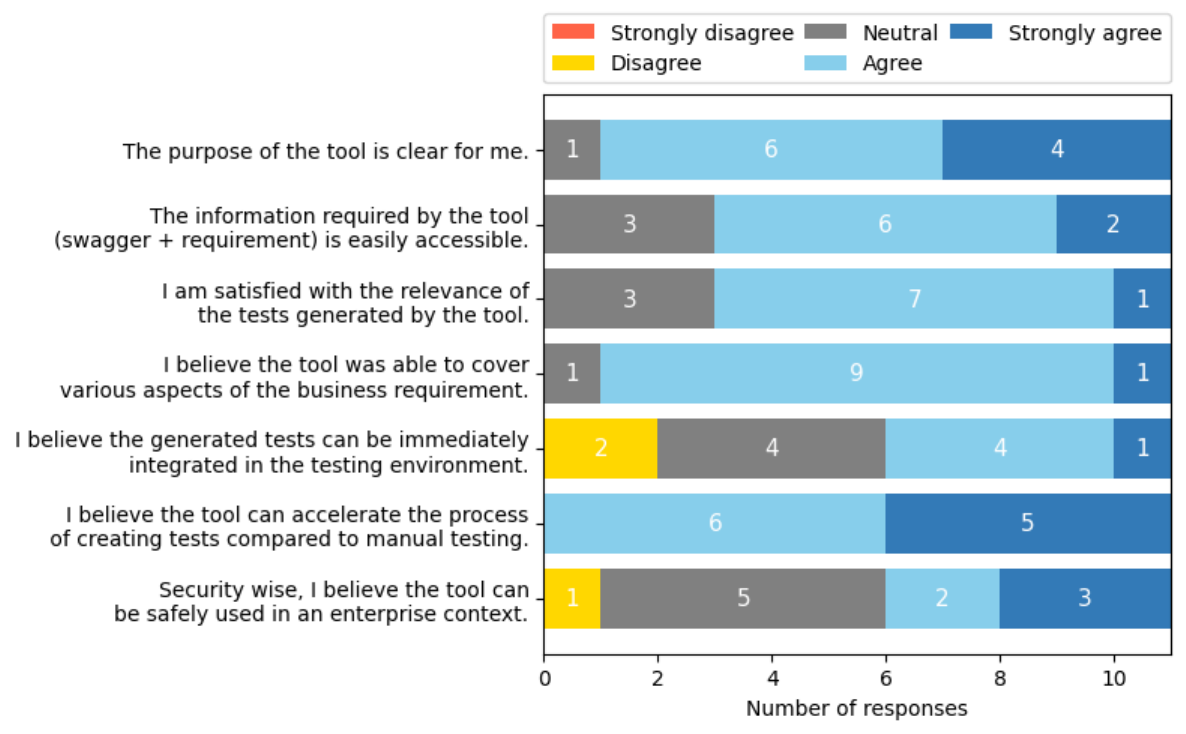}}
\caption{Opinions on \textit{APITestGenie}-generated tests from 11 practitioners with over five years of domain experience, collected anonymously after a workshop.}
\label{fig:believes}
\vspace*{-10pt}
\end{figure}

We conducted a hands-on \textit{APITestGenie} workshop with technical staff from our industrial partner in the automotive domain, who provided informed feedback on the relevance and completeness of the generated tests. During the workshop, we created four unique test scripts, demonstrating the tool's ability to handle progressively more complex requirements and APIs. Out of 15 participants, 11 responded to an anonymous survey, the results of which are shown in Figure~\ref{fig:believes}.

Overall, \textit{APITestGenie} was well received, with most participants finding the tests complete and relevant and expressing interest in daily use. Some participants noted the need for improvements in data security and in integration with existing testing environments.

All respondents agreed or strongly agreed that \textit{APITestGenie} can accelerate the test generation process compared to a manual approach. To better understand the potential gains, we conducted follow-up discussions with several workshop participants, who noted that manually scripting multi-endpoint API integration tests typically takes 20--30 minutes per case. 
In contrast, \textit{APITestGenie} reduces this to approximately two minutes per test, with an additional 2--5 minutes for manual review, yielding a 4--5$\times$ reduction in effort.  

These results suggest substantial return on investment in large-scale enterprise projects, particularly when paired with strong documentation and integration into existing workflows. By automating boilerplate test creation, \textit{APITestGenie} enables developers and QA engineers to focus on higher-value tasks such as handling edge cases and improving robustness.

\subsection{Answers to Research Questions}

\textbf{RQ1:} 
In our experiments, \textit{APITestGenie} effectively generated executable API test scripts from business requirements and API specifications. 
Across 75 generation attempts, 69.3\% of the scripts were syntactically and semantically valid, and at least one valid script was produced for 88.6\% of the business requirements within three attempts.  
The generation process was efficient, averaging 126 seconds per script, and required minimal human intervention.

\textbf{RQ2:} 
In our experiments, API complexity and level of detail in business requirements significantly influenced test generation effectiveness. 
Low-complexity APIs yielded more valid scripts (87.5\%) than high-complexity ones (60.8\%; p = 0.038), and requirements with concrete data achieved higher success rates (88.9\%) than those without (63.2\%; p = 0.039). 
Documentation detail showed no significant effect (p = 0.57). 
Overall, structural simplicity and explicit data examples were stronger predictors of successful automated test generation than documentation thoroughness.

\textbf{RQ3:} 
Industry practitioners perceived \textit{APITestGenie} as both useful and adoptable for API test generation. 
All survey respondents agreed that it accelerates test creation compared to manual scripting, and most found the generated tests relevant and complete. 
Participants expressed interest in using the tool in daily practice, noting that it could substantially reduce repetitive work. 
However, they also identified adoption barriers, including data security concerns when using external LLMs and the need for seamless integration with existing testing workflows. 
Overall, the feedback indicates a positive perception of \textit{APITestGenie}’s usefulness and practicality, contingent on addressing these integration and security requirements.

    % \item Results may vary with future LLM updates or API changes. To support reproducibility in a double-blind setting, we include anonymized artifacts (dataset, prompts, and generated scripts) as supplementary material; de-anonymized resources will be made available upon acceptance.

\subsection{Limitations and Threats to Validity}

This study has several limitations and potential threats to validity that should be considered when interpreting the results.

\subsubsection{Internal validity} 
Manual inspection, while systematic, introduces some subjectivity in assessing semantic validity. 
The analysis of influencing factors relied on naturally occurring variations in the available APIs and business requirements, which were not experimentally controlled, limiting causal inference. 
Nevertheless, this reflects the realities of industrial settings, where variables such as API complexity and documentation detail cannot be artificially balanced. 
To enhance internal validity, we applied consistent evaluation criteria, performed repeated generation attempts, and cross-checked ambiguous assessments among the authors. 
These measures improve reliability while acknowledging the inherent limitations of an observational design.

\subsubsection{External validity} 
Our evaluation involved ten APIs, eight of which came from a single industrial partner in the automotive domain. Consequently, results may not generalize to APIs in other domains, with different design styles or documentation practices. 
However, using industrial-grade APIs and experienced practitioners increased the practical validity of the study and provided realistic insights into tool applicability in production-like environments. 
User feedback was collected from 11 practitioners with over five years of experience, offering informed perspectives but limiting broader generalizability. 
The study compared \textit{APITestGenie} to manual test creation but not to other automated testing tools, as most existing approaches rely on API specifications, models, or execution traces rather than natural-language business requirements. 
Future studies should replicate the evaluation across additional domains and include comparative experiments with alternative automated approaches as they emerge to better position \textit{APITestGenie} within the broader landscape.

\subsubsection{Construct validity} 
The evaluation metrics captured generation effectiveness through the proportion of valid scripts and the number of business requirements with at least one valid script. 
While informative, the latter may overestimate practical usability, due to the manual review effort. 
On the other hand, the metrics do not reflect adaptive improvements that could arise if feedback from failed executions were used in subsequent prompts. 
Although the \textit{Test Improvement} flow in \textit{APITestGenie} supports such refinement, it was intentionally disabled to ensure comparability across the three independent attempts per requirement. 
Future work could enable this flow or add automatic regeneration after syntax or runtime errors to assess the benefits of self-correction and reduce manual review effort.

\section{Conclusion}
\label{sec:conclusion}

In this work, we presented \textit{APITestGenie}, an AI-driven tool for generating executable API test scripts from business requirements and OpenAPI specifications. Our approach leverages LLMs and RAG to scale testing in systems where traditional automation methods fall short, especially in complex, multi-endpoint integration scenarios.

Evaluated in collaboration with an industrial partner in the automotive domain, \textit{APITestGenie} demonstrated strong results: up to 89\% script validity with minimal manual intervention, and generation times fast enough to support agile testing workflows. The tool was well received by practitioners, who reported productivity gains and improved requirement traceability.

Importantly, our findings highlight that successful industrial adoption of GenAI tools depends not only on model performance but also on API complexity, documentation quality, clarity of business requirements, and seamless integration into development pipelines. While human validation remains necessary for critical testing paths, \textit{APITestGenie} already offers a valuable assistant to accelerate QA cycles.

Our evaluation targets the generation of integration tests from business requirements, which is distinct from most related tools that focus primarily on single-endpoint testing or rely exclusively on API specifications. By leveraging business requirements as input, \textit{APITestGenie} generates executable tests that simulate realistic, multi-endpoint scenarios, providing broader validation of service interactions and data flow. Although this focus means our results are not directly comparable with those of existing tools, \textit{APITestGenie} complements prior approaches by enabling comprehensive system-level testing that aligns more closely with actual business needs, as demonstrated in the comparison conducted with EvoMaster for one of the public APIs tested.

Looking forward, we plan to integrate the tool into CI/CD pipelines, address enterprise security considerations for LLM use, explore agent-based extensions, and conduct further evaluation studies. We believe that \textit{APITestGenie} represents a practical and scalable pathway to GenAI-enabled testing in real-world software engineering.

\section{Data and Code Availability}
\label{sec:data}

To support reproducibility and future research, whilst complying with the double-blind review policy, an anonymized repository containing a simplified version of the source code, example test scripts, and representative prompts used in this study is available at the folloinw url:
{\url{https://github.com/Andrepereira2001/ApiTestGenie}}.

\bibliographystyle{ACM-Reference-Format}
\bibliography{myrefs.bib}

\appendix

\section{Example of Test Script}

Listing~\ref{lst:testScript} presents an example of a generated test script interacting with the \texttt{/Post} and \texttt{/Get} endpoints to create a pet and retrieve it by ID. The execution of the test exposes a defect in the PetShop API, as the retrieval step fails.

%  caption={Example of a generated test script that interacts with two endpoints, \texttt{/Post} and \texttt{/Get}, to create a pet and subsequently retrieve it by ID. Upon execution, the retrieval step fails due to a defect in the PetShop API.},

\begin{lstlisting}[style=ts,
caption={Example of a generated test script interacting with the \texttt{/Post} and \texttt{/Get} endpoints.},
  label={lst:testScript}]
import axios from 'axios';

describe('Adding a New Pet to the Store', () => {
  const baseUrl = 'https://petstore.swagger.io/v2';
  let createdPetId;

  test('should create a unique pet', async () => {
    const newPet = {
      name: 'Fluffy',
      photoUrls: ['http://example.com/photo1'],
      category: { id: 1, name: 'Dogs' },
      tags: [{ id: 1, name: 'Friendly' }],
      status: 'available'
    };

    const response = await axios.post(`${baseUrl}/pet`, newPet, {
      headers: {
        'Content-Type': 'application/json'
      }
    });

    // Assuming the response contains the created Pet object
    expect(response.status).toBe(200);
    expect(response.data).toHaveProperty('id');
    createdPetId = response.data.id; // Store the ID for further use
    expect(response.data.name).toBe('Fluffy');
  });

  test('should retrieve the created pet by ID', async () => {
    if (!createdPetId) {
      throw new Error('Failed to store the created pet ID');
    }

    const response = await axios.get(`${baseUrl}/pet/${createdPetId}`, {
      headers: {
        'Content-Type': 'application/json'
      }
    });

    expect(response.status).toBe(200);
    expect(response.data.id).toBe(createdPetId);
    expect(response.data.name).toBe('Fluffy');
  });
});
    \end{lstlisting}

\end{document}